# Spin Texture in Quantum Point Contacts in the Presence of Lateral Spin Orbit Coupling

J. Wan[1], M. Cahay[1,*], P. Debray[2], and R.S. Newrock[2]

[1] Department of Electrical Engineering, University of Cincinnati, Cincinnati OH 45221, USA
[2] Department of Physics, University of Cincinnati, Cincinnati OH 45221, USA
*Corresponding author. Email: Marc.Cahay@uc.edu

## Abstract

A non-equilibrium Green's function formalism is used to study in detail the ballistic conductance of asymmetrically biased side-gated quantum point contacts (QPCs) in the presence of lateral spin-orbit coupling and electron-electron interaction for a wide range of QPC dimensions and gate bias voltage. Various conductance anomalies are predicted below the first quantized conductance plateau ($G_0 = 2e^2/h$) which occur due to spontaneous spin polarization in the narrowest portion of the QPC. The number of observed conductance anomalies increases with increasing aspect ratio (length/width) of the QPC constriction. These anomalies are fingerprints of spin textures in the narrow portion of the QPC.

# I. Introduction

Since spin-orbit coupling (SOC) couples the electron orbital motion to its spin, it is expected to play a role in triggering spin polarization in quantum point contacts (QPCs) made of materials with a strong intrinsic SOC without ferromagnetic contacts and applied magnetic fields. Considerable research has been done to use the Rashba spin-orbit coupling (RSOC)[1,2] for achieving this objective. Despite this, there has been no report of success in experimentally achieving spin polarization by RSOC. We have recently used lateral SOC (LSOC) resulting from the lateral in-plane electric field of the confining potential of a side-gated QPC to create strongly spin-polarized current by purely electrical means[3] *in the absence of* applied magnetic field. Using NEGF analysis of a small model QPC[3], three ingredients were found to be essential to generate the strong spin polarization: an asymmetric lateral confinement, a LSOC induced by the lateral confining potential of the QPC, and a strong electron-electron (e-e) interaction. In a recent work[4] we modeled QPCs with size close to that of the experimental device. Using NEGF formalism we were able to reproduce the 0.5 $G_0$ plateau observed experimentally[3].

NEGF simulations provide a general approach for quantum transport far from equilibrium treating e-e interactions in a mean field framework[5] and have been used by several authors to study the conductance through QPCs[6-9]. We have earlier used this formalism to investigate the effects of LSOC[3,4] and found spin symmetric ground states at equilibrium with no applied drive bias voltage $V_{ds}$ between the source and drain. When a finite $V_{ds}$ is applied, a current flows through the QPC and the system passes into the non-equilibrium sate. The current flow breaks the time reversal symmetry. As a result, the spin symmetry is broken and a spin polarization develops in the channel of the QPC[10]. *This happens even in the absence of e-e interaction*[3]. *The breaking of spin symmetry in our model is a purely non-equilibrium effect*. It is similar to the Spin Hall Effect[11-14] which has also been modeled using the NEGF method[15]. For our QPCs with LSOC in the presence of a finite bias $V_{ds}$, there is no net spin polarization in the channel if the same bias is applied to the two side gates and the confining potential is symmetric. The application of asymmetric bias voltages to the two side gates makes the confining potential asymmetric. The resulting asymmetric LSOC triggers a net spin polarization which can reach near 100% in the presence of strong e-e interaction[3,4].

For more than a decade, there have been many experimental reports of anomalies appearing at non-integer values of the quantized conductance $G_0 = (2e^2/h)$ in the ballistic conductance of quantum point contacts (QPCs) based on GaAs. These include the observation of an anomalous plateau at $G \cong 0.5 G_0$[16-18] and the well-known "0.7 structure"[19]. In some of these cases the landscape of the QPC confining potential was found to play a crucial role[17], while in others the devices were created by multiple gates[16], making it difficult to have a clear idea of the symmetry of the QPC confining potential when the anomalies are observed. The 0.7 anomaly has been observed in GaAs QPCs with apparently symmetric confinement. This anomaly, however, is not a universal feature since it is not observed in all devices. Also, it can be made to appear or disappear by tuning the bias voltages of the QPC gates and hence by changing the details of the QPC lateral confining potential[20,21]. The role of asymmetric QPC potential confinement in the appearance of the above conductance anomalies cannot, therefore, be ruled out. The physical origin of the anomalies observed in the ballistic conductance of QPCs is still a highly debated issue.

In this work we present NEGF calculations of the conductance of symmetrically and asymmetrically biased side-gated QPCs in the presence of LSOC with and without e-e interaction. Numerical results are presented for asymmetrically confined QPCs with a wide range of device length, aspect ratio, and biasing parameters in the presence of e-e interaction. Many conductance anomalies are found at non-integer values of $G_0 = (2e^2/h)$ the number of which increases with increasing aspect ratio of the QPC. The occurrence of these anomalies is related to the existence of a plethora of spin textures in the narrowest part of the QPC which show either a Spin Hall regime or a spontaneous net spin polarization. We believe this is the first theoretical work that predicts spin textures as the possible common origin of the conductance anomalies observed in QPCs.

In section II, we present the results of extensive NEGF simulation results for a wide range of QPC dimensions and biasing parameters. A variety of conductance anomalies are observed at non-integer values of $G_0$, especially for QPCs with asymmetric confinement in the presence of e-e interaction. In section III, we show that the appearance of these anomalies is linked to the existence of spin textures in the narrowest part or the constriction of the QPC. Our results are complimentary to the work of Xu and coworkers[22,23], who studied the onset of spin textures in QPCs in the case of multimode transport in the presence of both Rashba and Dresselhaus spin orbit coupling. They showed that, for spin-polarized injection, the electron probability distribution has a zitterbewegung pattern in the QPC. For spin-unpolarized injection, the spin-dependent electron density shows spatially dependent stripes of spin accumulations. However, because their work did not include an asymmetric confining potential, no net polarization was found in the narrow portion of the QPC. Finally, in section

IV we present our concluding remarks.

## II. Numerical Results

The model QPC we have used is shown in Fig. 1, where the white region represents the QPC channel with openings at the ends. The gray area represents the etched isolation trenches that define the lithographic dimensions of the QPC constriction. The black strips show the four contact electrodes connected to the QPC device: source, drain and two side gates (SGs). Symmetric and asymmetric SG voltages can be applied. We consider the QPC in Fig. 1 to be made from a nominally symmetric InAs quantum well (QW). Spatial inversion asymmetry is therefore assumed to be negligible along the growth axis (*z* axis) of the QW and the corresponding Rashba spin-orbit interaction is neglected. The Dresselhaus spin-orbit interaction due to the bulk inversion asymmetry in the direction of current flow was also neglected. The only spin-orbit interaction considered is the LSOC due to the lateral confinement of the QPC channel, provided by the isolation trenches and the bias voltages of the side gates[3,4]. The free-electron Hamiltonian of the QPC is given by

$$H = H_0 + H_{SO}$$
$$H_0 = \frac{\hbar^2 (k_x^2 + k_y^2)}{2m^*} + U(x,y) \qquad (1)$$
$$H_{SO} = \beta \vec{\sigma} \cdot (\vec{k}_x \times \vec{\nabla} U(x,y))$$

In equation (1), $H_{SO}$ is the LSOC interaction term, $\beta$ the intrinsic SOC parameter, $\vec{\sigma}$ the vector of Pauli spin matrices, and $\vec{B}_{SO} = \beta(\vec{k}_x \times \vec{\nabla} U(x,y))$ is the effective magnetic field induced by LSOC. The low band gap semiconductor InAs has a large intrinsic SOC. The effective mass in the InAs channel was set equal to $m^* = 0.023 m_0$, where $m_0$ is the free electron mass. The 2DEG is assumed to be located in the (*x*, *y*) plane, *x* being the direction of current flow from source to drain and *y* the direction of transverse confinement of the channel. $U(x, y)$ is the confinement potential, which includes the potential introduced by gate voltages and the conduction band discontinuity at the InAs/air interface.

The conductance through the QPC was calculated using a NEGF method under the assumption of ballistic transport[4]. We used a Hartree-Fock approximation following Lassl et al.[9] to include the effects of e-e interaction in the QPC. More specifically, the e-e interaction was taken into account by considering a repulsive Coulomb contact potential, $V_{int}(x,y;x',y') = \gamma\, \delta(x-x')\, \delta(y-y')$, where $\gamma$ indicates the e-e interaction strength. As a result, an interaction self-energy, $\Sigma_{int}^{\sigma}(x, y)$, must be added to the Hamiltonian in Eq.(1). The parameters $\gamma$ and $\beta$ were set equal to $3.7(\hbar^2/2m^*)$ and $2.0 \times 10^{-18} m^2$, respectively. The value of $\gamma$ chosen is moderate for a 1D electron system and can be much higher for low electron density. The value of $\beta$ chosen is appropriate for InAs and its exact value is not of crucial importance[3]. For all simulations, $V_s = 0V$, $V_d = 0.3mV$. An asymmetry in the QPC potential confinement was introduced by taking $V_{sg1} = 0.2\ V + V_{sweep}$ and $V_{sg2} = -0.2\ V + V_{sweep}$ and the conductance of the constriction was studied as a function of the sweeping (or common mode) potential, $V_{sweep}$. At the interface between the rectangular region of size $w_2 \times l_2$ and vacuum, the conduction band discontinuities at the bottom and the top interface were modeled, respectively, as

$$\Delta E_c(y) = \frac{\Delta E_c}{2}\left[1 + \cos\frac{\pi}{d}\left(y - \frac{w_1 - w_2}{2}\right)\right] \quad (2)$$

and

$$\Delta E_c(y) = \frac{\Delta E_c}{2}\left[1 + \cos\frac{\pi}{d}\left(\frac{w_1 + w_2}{2} - y\right)\right] \quad (3)$$

to achieve a smooth conductance band change, where *d* was selected to be in the nm range to represent a gradual variation of the conduction band profile from the inside of the quantum wire to the vacuum region. A similar grading was also used along the walls going from the wider part of the channel to the central constriction of the QPC (Fig.1). This gradual change in $\Delta E_c(y)$ is responsible for the LSOC that triggers the spin polarization of the QPC in the presence of an asymmetry in $V_{sg1}$ and $V_{sg2}$. The parameter $d$ appearing in Eqns. (2) and (3) was set equal to 1.6 nm. Similar results were obtained when $\Delta E_c(y)$ is linearly changed at the interface and $d$ was set equal to 0.8 nm and 1.2 nm. The conductance of the QPC was then calculated using the NEGF with a non-uniform grid configuration containing more grid points at the interface of the QPC with vacuum. All calculations were performed at a temperature $T = 4.2\ K$.

**Conductance modulation as a function of length**

Figure 2 shows the conductance $G$ of asymmetrically biased ($\Delta V_{SG} = V_{sg1} - V_{sg2} = 0.4$ V) QPC of Fig. 1 as a function of sweeping voltage $V_{sweep}$ applied to both gates for a range of device dimensions in the presence of e-e interaction. Figures 2a and 2b show the conductance for different values of $l_2$ with $w_2 = 16$nm, $w_1 = 48$nm, and $l_1 = l_2 + 32$nm, with $l_2$ varying from 22 to 50 nm in steps of 2 nm. A wide plethora of conductance anomalies can be seen which depends strongly on the QPC length. In Fig. 2a, the first hint of an anomalous conductance plateau appears as a shoulder for $l_2 = 26$nm (curve 3), at aspect ratio of $l_2/w_2 = 1.625$. For $l_2 = 28$nm (curve 4), the small shoulder has grown into an anomalous plateau slightly below $0.7G_0$, accompanied by the onset of a negative differential resistance (NDR). The latter becomes more pronounced as the length or aspect ratio $l_2/w_2$ of the QC increases. As $l_2$ keeps increasing, there is a pinning of the conductance curve around $0.5\ G_0$ prior to the onset of the NDR. In Fig.2b for $l_2$ = 42 nm (curve 11), the conductance curve actually shows two simultaneous anomalies, one around 0.5 $G_0$ and the other around 0.75 $G_0$. For $l_2$ = 50nm (curve 15), the small NDR which originally appeared for $l_2$ = 42 nm (curve 11) has grown into a second NDR with nearly the same peak and valley location on the conductance axis as the first NDR located past the 0.5 $G_0$ plateau.

**Conductance modulation as a function of width**

Figure 2c shows the conductance for different widths of the QPC with the other parameters selected as follows, $w_1$ = 3$w_2$, $l_2$ = 46nm, $l_1$ = $l_2$ + 2$w_2$, with $w_2$ varying from 16 to 28 nm in steps of 2 nm. The results show a trend similar to the one displayed in Figs. 2a and 2b. First, the onset of a conductance anomaly occurs for $w_2$=28nm with an aspect ratio of $l_2/w_2$ = 1.64 (curve 6), which develops into a plateau that eventually saturates at 0.5 $G_0$. The latter is followed by two NDRs with increasing peak to valley ratio as the aspect ratio of the QPC increases ($l_2/w_2$ = 2.875 for curve 1).

### III. Discussion

To understand the appearance of the anomalous conductance plateaus, we calculated the conductance curves for $l_2$=22, 28, 42, and 50 nm with and without an asymmetric bias between the two SGs and with and without the e-e interaction. These are shown in Fig. 3. The panels labeled (a), (b), (c), and (d) correspond, respectively, to $l_2$=22, 28, 42, and 50 nm. Figure 3 reveals the following important features: First, there is little difference between the conductance plots of symmetric and asymmetric QPC confinements if the effect of e-e interaction is neglected. Second, the inclusion of e-e interaction shifts the conductance curves to the right, i.e., the pinch-off occurs at less negative values of $V_{sweep}$. This means that the overall barrier in the central portion of the QPC moves upward in energy as the effect of e-e interaction is taken into account. Third, for asymmetric QPC confinement, the inclusion of e-e interaction, brings the conductance curve down compared to the case when e-e interaction is neglected. This vertical shift is stronger for larger aspect ratios of the QPC. Figure 3 clearly shows that the conductance anomalies observed when e-e interaction is included are evolutions of the conductance oscillations in the QPC calculated in the single electron picture. For instance, for $l_2 = 50$ nm, the double NDR appearing on the 0.5 $G_0$ plateau is a vertical shift of the first two oscillations of the conductance curve calculated when e-e interaction is neglected. A similar behavior is seen for the corresponding curves for $l_2 = 28$ and 42 nm. In this case, the period of the oscillations calculated without e-e interaction is larger because the spacing between the quasi-bound state energies (Ramsauer oscillations) above the average barrier in the QPC in the direction of current flow is larger. The observation of NDR accompanying anomalous conductance plateaus has been reported both experimentally[22] and theoretically[23]. The NDR peak-to-valley ratio observed experimentally is much smaller than the theoretical values. This is due to the fact that the model QPC used for simulation is narrower in width than the experimental device, for which the subband separation is smaller and therefore thermal smearing tends to wash out the NDR.

The conductance anomalies illustrated in Fig.2 are all linked to non-zero values of the conductance spin polarization $\alpha = (G_\uparrow - G_\downarrow)/(G_\uparrow + G_\downarrow)$, where $G_\uparrow$ and $G_\downarrow$ are, respectively, the conductance associated with the spin-up and spin-down electrons[4]. We illustrate this by considering the conductance curves as a function of $V_{sweep}$ shown in Fig.4 of asymmetric

QPCs with e-e interaction for $l_2$ = 22, 28, 42, and 50 nm. Figures 5-8 show two-dimensional contours of the spin density profiles $n_\uparrow(x,y) - n_\downarrow(x,y)$ as a function of $V_{sweep}$ associated with the conductance plots shown in Fig.4. A Spin Hall state regime in the central constriction of the QPC (in which spin of opposite polarities appear on opposite sides of the channel) is clearly seen near the threshold for conduction and near the first conductance step. For QPC with increasing aspect ratio, the spin texture is quite different near the anomalous conductance steps, as discussed below.

**Case $l_2$ = 22nm**

In this case, there is no anomalous plateau in the conductance and the Spin Hall regime prevails over the entire range of $V_{sweep}$, as shown in Fig.5. The spin density on both sides of the channel increases by about a factor of four as $V_{sweep}$ goes from – 150 mV to 300 mV when the first integer conductance plateau is reached. This is due to the second bound state associated with the lateral confinement in the QPC which favors electron accumulation on both sides of the channel.

**Case $l_2$ = 28nm**

For $V_{sweep}$ below - 90 mV, a Spin Hall texture exists in and around the central part of the QPC (Fig.6a). For $V_{sweep}$ between -90 and 210 mV, the spin texture is characterized by a sharply defined hump located at the center of the QPC. The regime around $V_{sweep}$= 0 mV shows a strongly spin-up polarized hump (Fig. 6b) and corresponds to onset of the anomalous conductance plateau observed slightly below 0.7 $G_0$ (Fig. 4). As $V_{sweep}$ increases, the hump slowly disappears (Fig. 6c) and beyond $V_{sweep}$ = 240 mV, the spin texture switches back to the Spin Hall regime (Fig.6d).

**Case $l_2$ = 42nm**

For $V_{sweep}$ below - 150 mV, the spin texture is reminiscent of a Spin Hall regime (Fig.7a). As the threshold for the anomalous 0.5 $G_0$ plateau is reached ($V_{sweep}$ = - 90mV, Fig.7b), the spin density in the channel as increased by several orders of magnitude and is mostly due to spin-up electrons. The spin texture consists of a large hump located at the center of the QPC. As $V_{sweep}$ increases, this single bump evolves into a spin texture composed of two coexisting lobes (one for majority and the other for minority spin) in the direction of current flow (Fig.7c). This occurs at the onset of the plateau near 0.75 $G_0$ (Fig.4). The two lobes are present over the voltage range from 60 to 90 mV. Beyond this second plateau, the spin texture returns to a Spin Hall regime (Fig.7d) as the first integer conductance step is reached (Fig.4).

**Case $l_2$ = 50nm**

For $V_{sweep}$ close to the threshold ($\leq$ -150 mV) the spin texture shows a Spin Hall regime (Fig.8a). After the onset of the 0.5 $G_0$ plateau and in the range of the two NDRs, the spin texture has a single hump located at the center of the QPC (Fig.8b,c). Past the NDR range, the spin texture is composed of two coexisting lobes (one for majority and the other for minority spin) (Fig.8d), as in the case $l_2$ = 42nm. In this case, no anomalous conductance step can be seen around 0.75 $G_0$. For $V_{sweep} \geq$ 180 mV, as the conductance approaches the integral value $G_0$, the spin texture reverts back to the spin Hall pattern (Fig.8e,f).

## IV. Conclusions

The physical origin of the anomalous conductance anomalies is still a highly debated issue. Majority of the theoretical models link them to spontaneous spin polarization in the QPC[20,24-26]. Our earlier NEGF work[3,4] has shown that a strong spin-orbit coupling is not essential for generating strong spontaneous spin polarization. Needed is an initial spin imbalance, howsoever small, that is enhanced by strong e-e interaction. This initial spin imbalance can be provided by asymmetric LSOC. Spin textures and spontaneous spin polarization discussed here may therefore occur in semiconductor QPCs, including those based on GaAs and other semiconductors with weak spin-orbit coupling. We believe the common origin of the anomalies observed in the ballistic conductance of QPCs is the presence of spin textures that result from a combination of large device aspect ratio, asymmetric LSOC, and the presence of strong e-e interaction. We hope this work will provide guidelines to device physicists in their effort to develop all-electric spin devices.

**Acknowledgment**

We thank Supriyo Datta and Avik Ghosh for insightful discussions. This work was supported by NSF award ECCS 0725404.

**Figure Captions**

**Fig.1.** Schematic illustration of the QPC geometrical layout used in the numerical simulations. In all simulations, $V_s$ = 0V, $V_d$ = 0.3mV.

**Fig.2.** Conductance, $G$, of asymmetrically biased QPC as a function of $V_{sweep}$. The potential on the two SGs are $V_{sg1}$ = 0.2V + $V_{sweep}$ and $V_{sg2}$ = -0.2V + $V_{sweep}$. The temperature is set equal to 4.2K. In panels **a** and **b,** $l_1$ = $l_2$ + 32nm, $w_2$ = 16nm, and $w_1$ = 48nm. The Fermi energy was taken equal to 106.3 meV in the source contact and 106 meV in the drain contact, ensuring single-mode transport through the QPC. In **a**, the curves labeled 1 through 11 correspond to $l_2$ = 22, 24, 26, 28, 30, 32, 34, 36, 38, 40, and 42 nm, respectively. In **b**, the curves labeled 11 through 15 correspond to $l_2$ = 42, 44, 46, 48, 50 nm, respectively. In **c**, $l_1$= $l_2$ + 32nm, $l_2$ = 46nm, and $w_1$ = 3 $w_2$. The curves labeled 1 through 7 correspond to $w_2$ = 16, 18, 20, 22, 24, 26, and 28 nm. For these curves, the Fermi level in the source contact was set equal to 106.3 meV, 81.8meV, 64.8 meV, 52.6meV, 43.5meV, 36.6meV and 31.2meV, respectively. In all plots, $V_{ds}$ = 0.3mV, T = 4.2K, $\gamma$ = 3.7 ($\hbar^2/2m^*$), and $\beta$ = 200 Å$^2$.

**Fig.3.** Conductance, $G$, of QPC as a function of $V_{sweep}$ with and without the effect of the e-e interaction for asymmetric and symmetric biasing of the QPC. For the symmetric side-gate voltages, $V_{sg1}$= $V_{sg2}$ = $V_{sweep}$; for the asymmetric side-gate voltages, $V_{sg1}$ = 0.2V + $V_{sweep}$ and $V_{sg2}$ = -0.2V + $V_{sweep}$. In all plots, the Fermi energy was taken equal to 106.3 meV in the source contact and 106 meV in the drain contact, ensuring single-mode transport through the QPC. In panels (a), (b), (c), and (d), $l_2$ is set equal to 22, 28, 42, and 50 nm, respectively. In each panel, $l_1 = l_2 + 32$nm, $w_2 = 16$nm, and $w_1 = 48$nm. The plots are characteristic of QPCs with small, intermediate and large aspect ratio. The solid blue curve corresponds to the case of asymmetric $V_{sg}$ without e-e interaction, the green open circles to the case of symmetric $V_{sg}$ without e-e interaction, the solid black curve to the case of asymmetric $V_{sg}$ with e-e interaction, and the solid red curve to the case of symmetric $V_{sg}$ with e-e interaction. In all plots, $V_{ds} = 0.3$mV, T = 4.2K, $\gamma = 3.7$ in units of $\hbar^2/2m^*$, and $\beta = 200$ Å$^2$.

**Fig.4.** Conductance, $G$, of asymmetric QPC with e-e interaction as a function of $V_{sweep}$ for $l_2 = $ 22, 28, 42, and 50 nm, reproduced from Fig.3.

**Fig.5.** Two-dimensional contour plots of spin density for discrete values of $V_{sweep}$ associated with the conductance plot for $l_2 = 22$nm (Fig.4). **a,** $V_{sweep} = -150$ mV. **b,** $V_{sweep} = 0$ mV. **c,** $V_{sweep} = 150$ mV. **d,** $V_{sweep} = 300$ mV.

**Fig.6.** Two-dimensional contour plots of spin density as a function of $V_{sweep}$ associated with the conductance plot for $l_2 = 28$nm (Fig.4). **a,** $V_{sweep} = -150$ mV. **b,** $V_{sweep} = -0$ mV. **c,** $V_{sweep} = 210$ mV. **d,** $V_{sweep} = 300$ mV.

**Fig.7.** Two-dimensional contour plots of spin density as a function of $V_{sweep}$ associated with the conductance plot for $l_2 = 42$nm (Fig.4). **a,** $V_{sweep} = -150$ mV. **b,** $V_{sweep} = -90$ mV. **c,** $V_{sweep} = 60$ mV. **d,** $V_{sweep} = 300$ mV.

**Fig.8.** Two-dimensional contour plots of spin density as a function of $V_{sweep}$ associated with the conductance plot for $l_2$ = 50nm (Fig.4). **a,** $V_{sweep}$ = - 150 mV. **b,** $V_{sweep}$ = -90 mV. **c,** $V_{sweep}$ = 0 mV. **d,** $V_{sweep}$ = 30 mV. **e,** $V_{sweep}$ = 210 mV. **f,** $V_{sweep}$ = 300 mV.

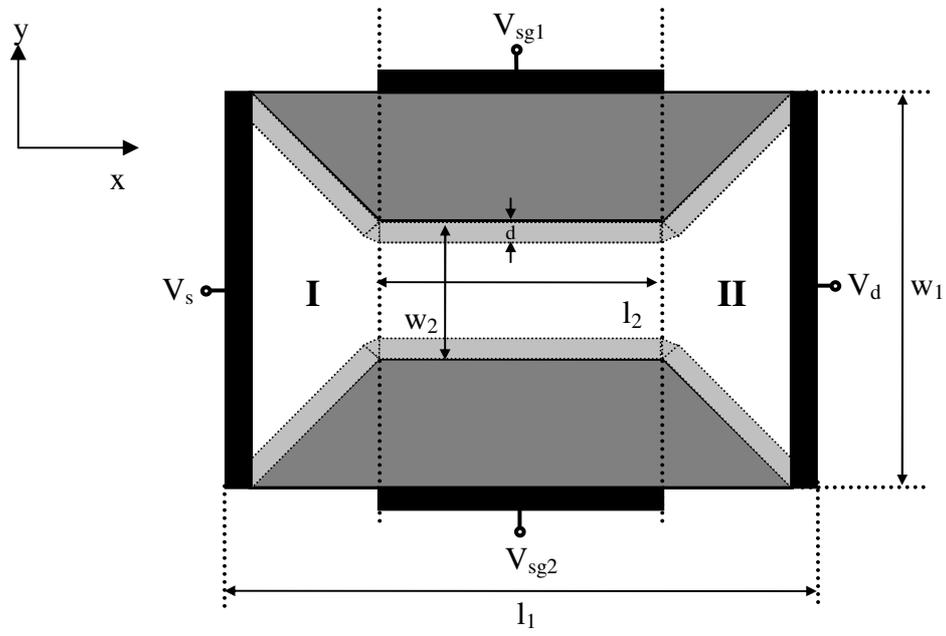

**Fig.1** (J. Wan et al.)

**a**

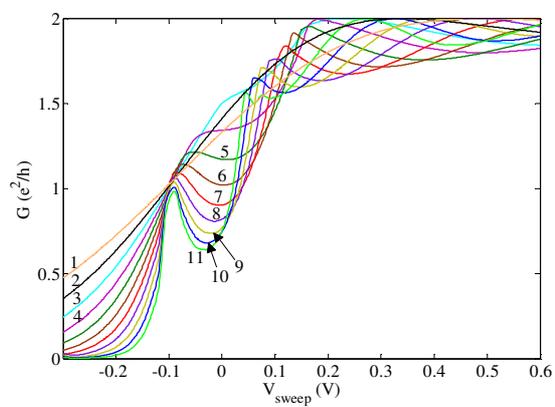

**b**

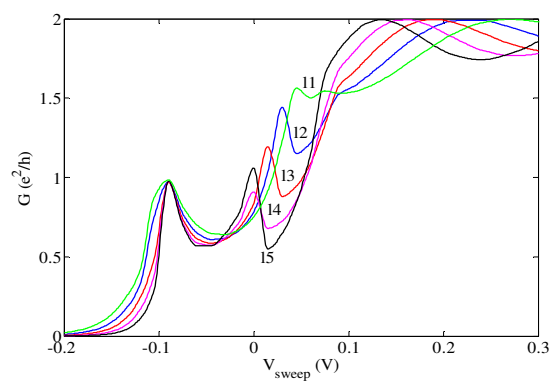

**c**

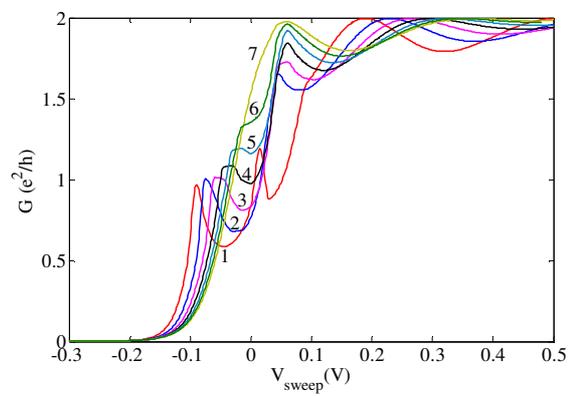

**Fig.2** (J. Wan et al.)

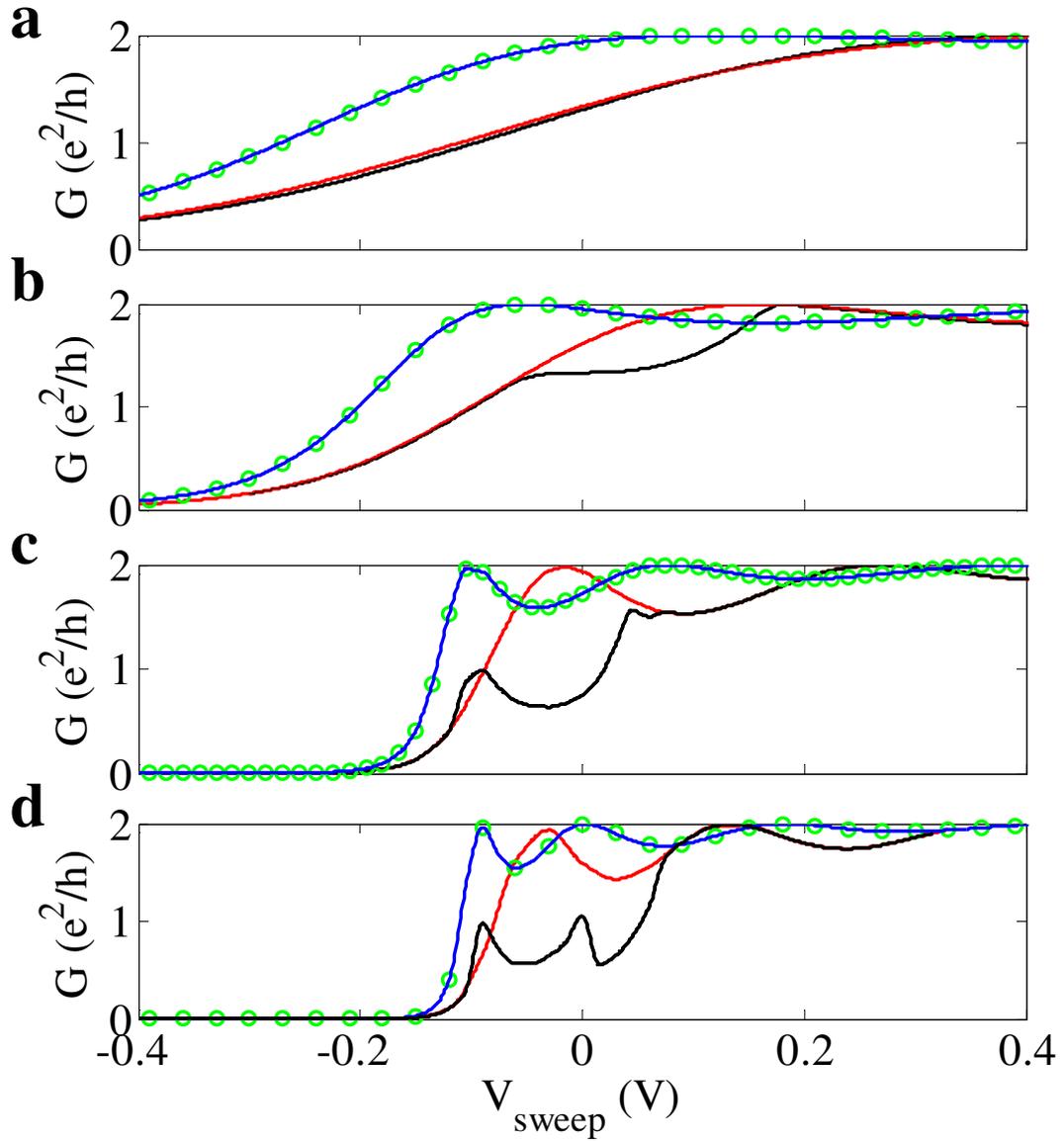

**Fig.3** (J. Wan et al.)
20

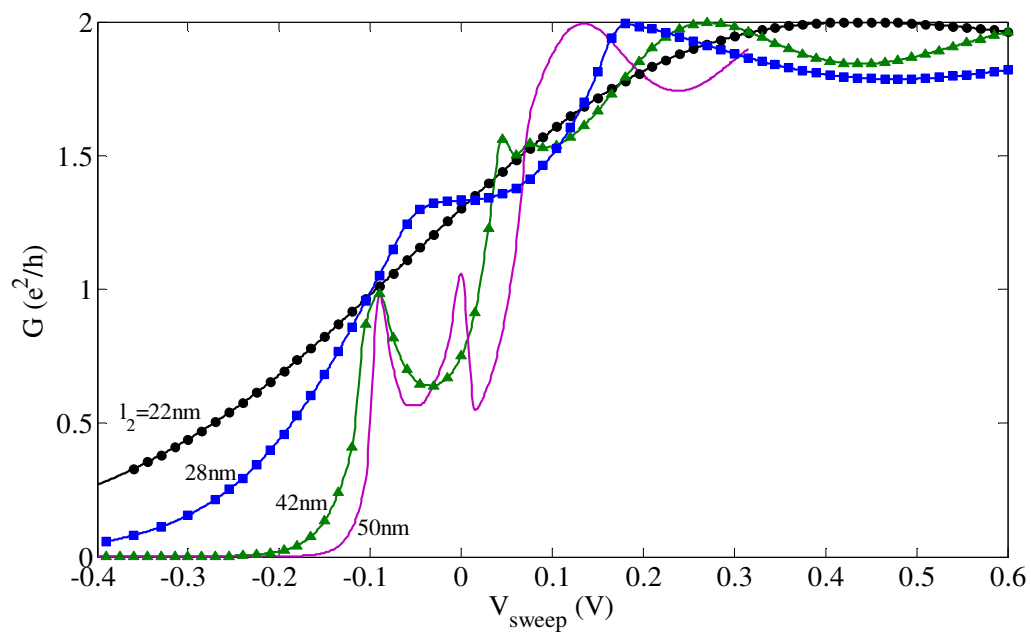

**Fig.4** (J. Wan et al)

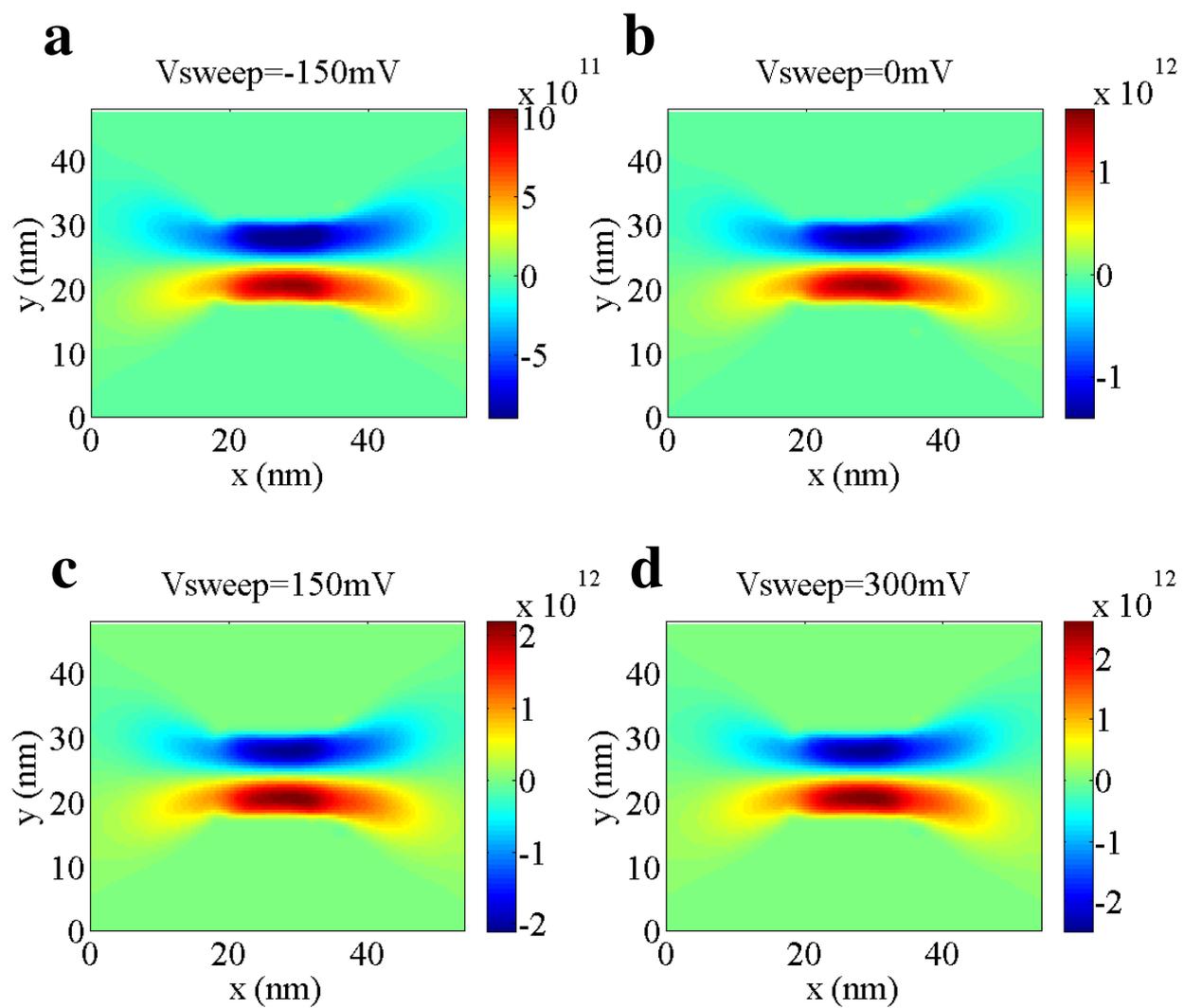

**Fig.5** (J. Wan et al.)

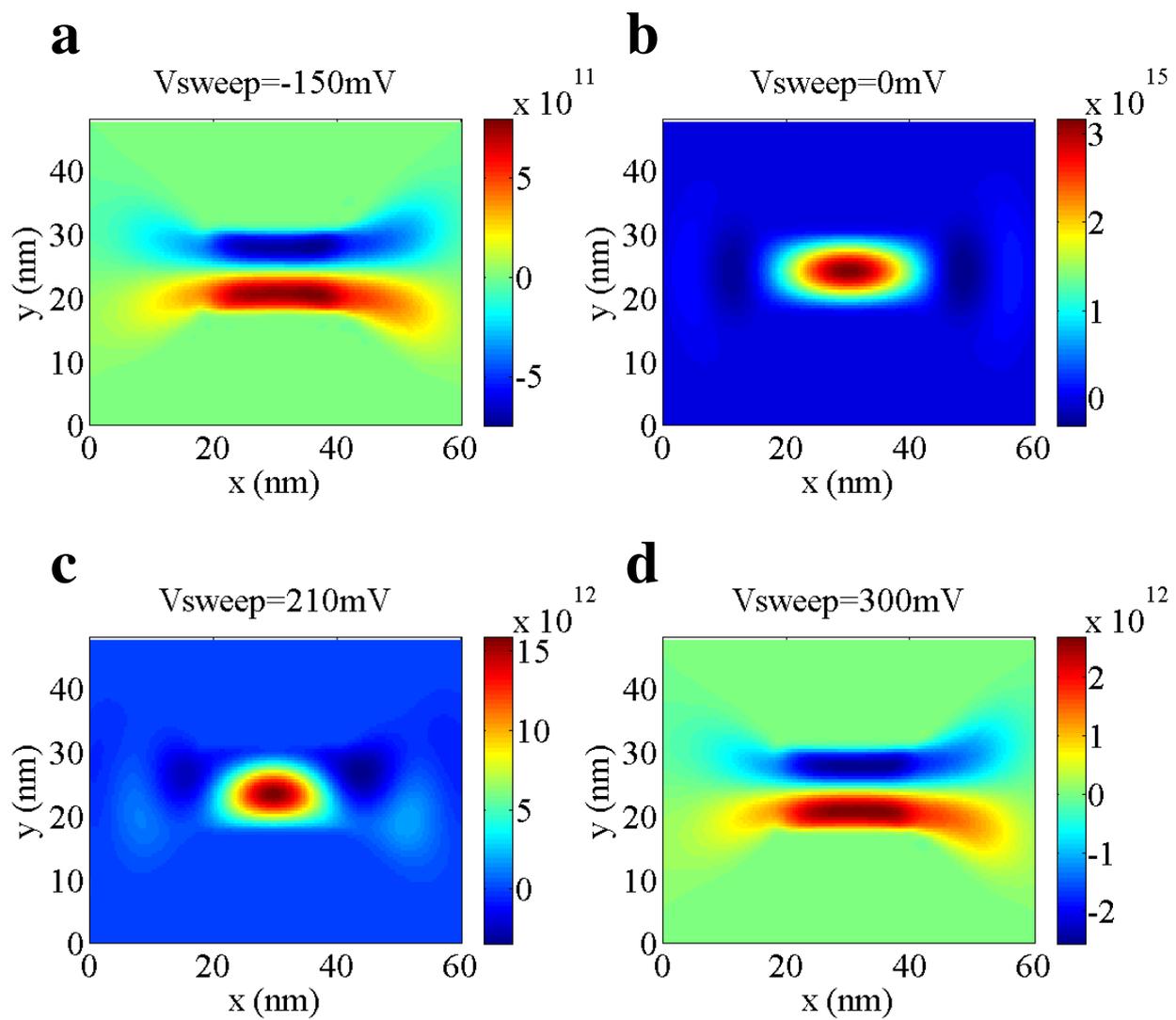

**Fig.6** (J. Wan et al.)

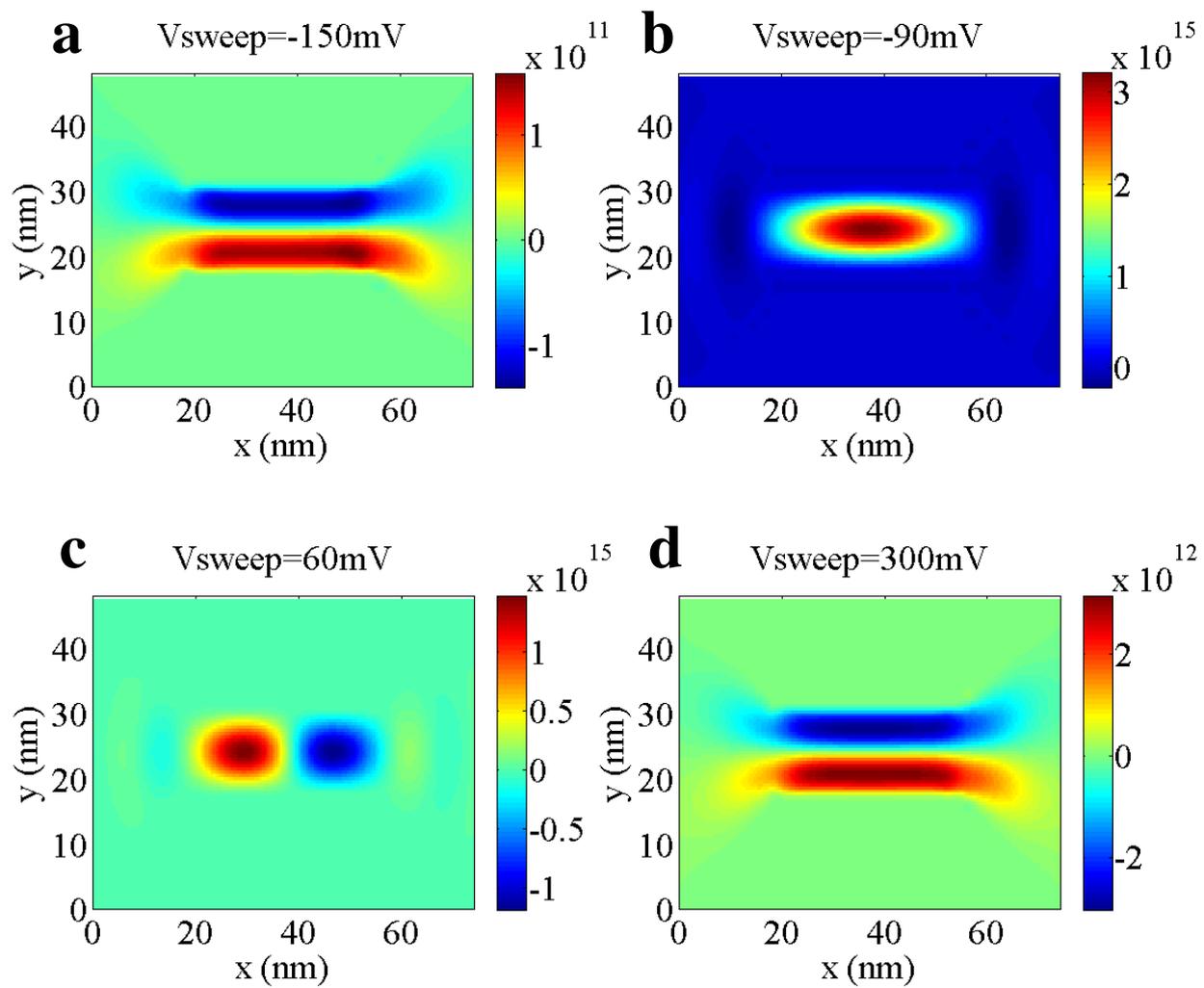

**Fig.7** (J. Wan et al.)

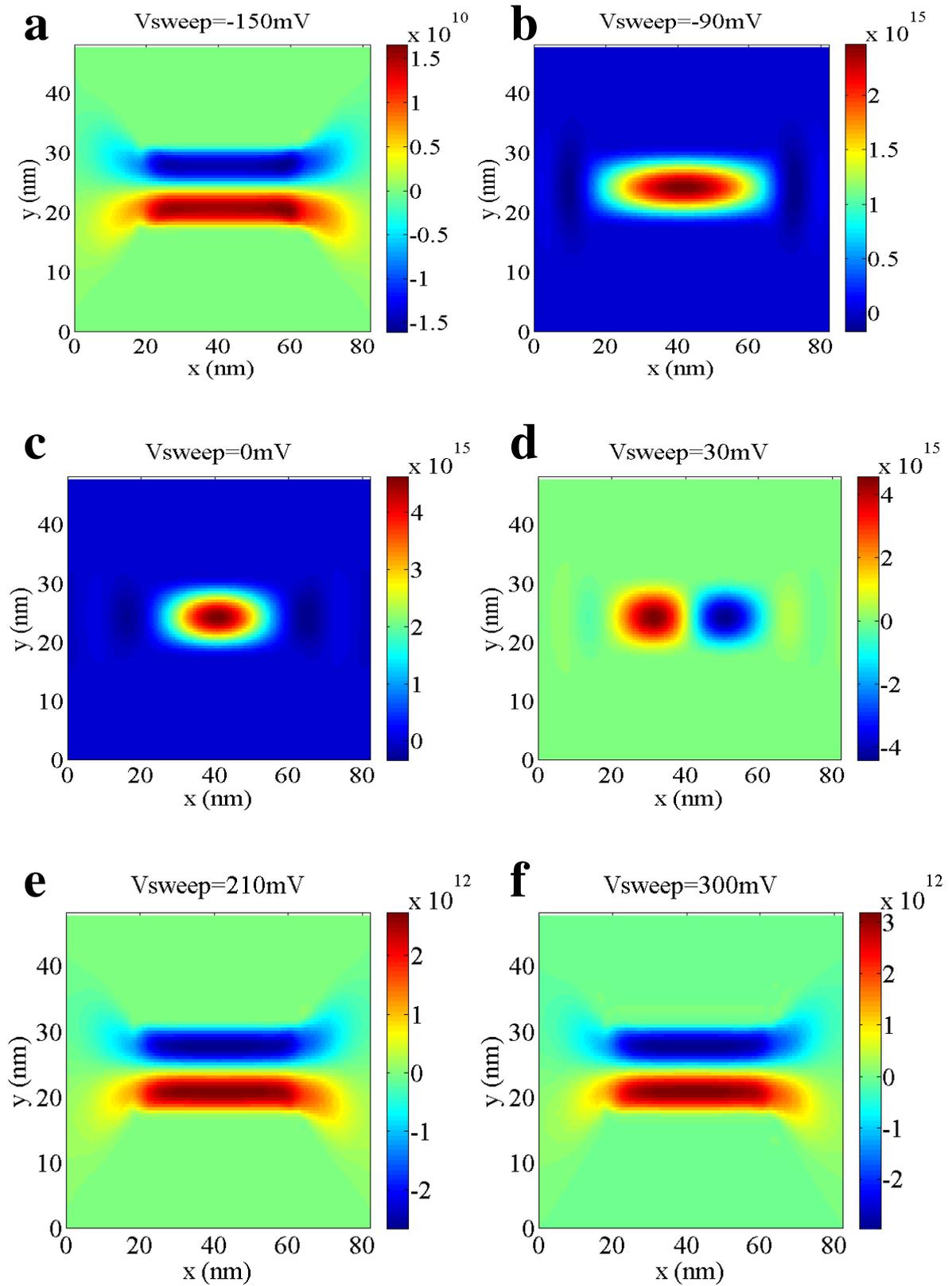

**Fig.8** (J. Wan et al)